# Modular AWG-based Optical Shuffle Network

**Jingjie Ding, Tong Ye, Tony T. Lee, and Weisheng Hu**
*State Key Laboratory of Advanced Optical Communication Systems and Networks, Department of Electronic Engineering
Shanghai Jiao Tong University, 800 Dongchuan Rd, Shanghai 200240, China
E-mail: {mrdingjay, yetong, ttlee}@sjtu.edu.cn*

**Abstract:** This paper proposes an arrayed-waveguide grating (AWG) based wavelength-division-multiplexing (WDM) shuffle network. Compared with previous optical shuffle networks, our proposal is compact, easy to implement, highly scalable, and cost effective.

## 1. Introduction

The shuffle network has widespread application in interconnection networks such as parallel processing operations [1] and switching networks [2]. Optical technology has been considered as a promising candidate to implement shuffle network [3], since optical interconnections have the advantages of low power consumption and high transmission bit-rates. Several kinds of optical shuffle networks have been demonstrated based on free space optics (FSO). In [4], an optical shuffle network was achieved by spatially interleaving two copies of the input light signal. This scheme requires the accurate adjustment of lenses and polarizing beam splitters. The optical folded shuffle proposed in [5] relies on a complicated lens array. In [6, 7], two kinds of optical shuffle are proposed by applying transceivers to emit and receive the light and using panels to reflect the light, where transceivers are active components and algorithms for calculating the spatial angle of the transceivers are required. Thus, the existing FSO-based shuffle networks are costly and difficult to adjust and not suitable for practical applications.

In this paper, we propose a kind of AWG-based shuffle network, which processes several attractive features as follows. Firstly, this scheme is compact and very easy to implement, because it only consists of AWGs interconnected by fiber links. Secondly, the AWG-based shuffle network is highly scalable, since large-scale optical shuffle can be constructed from a set of AWGs with small port count. Third, it is cost effective, for all the components employed have already been commercial.

## 2. Single AWG-based Shuffle Network

In this section, we firstly demonstrate that a single AWG can fulfill the function of a shuffle network using an example. An $N \times N$ classical shuffle network, denoted by $S(g, l)$, is illustrated in Fig. 1(a), where $N$ inputs and $N$ outputs are evenly divided into $g = 3$ groups and $l = 6$ groups. Each input is labelled by two-bit address $x_2 x_1$, while each output is labelled by $z_2 z_1$, where $x_2, z_1 = 0,1,\cdots, g-1$ and $x_1, z_2 = 0,1,\cdots, l-1$. The shuffle network can achieve left shift of port address. As an example, input $x_2 x_1$ in Fig. 1(a) is connected with output $z_2 z_1 = x_1 x_2$.

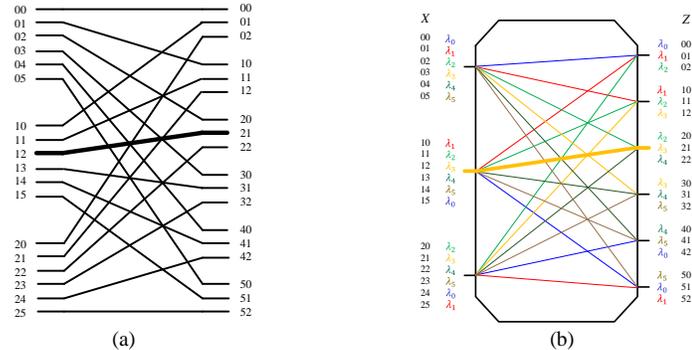

Fig. 1. The equivalence between (a) a $18 \times 18$ shuffle network and $S(3,6)$ and (b) a $3 \times 6$ AWG

A $g \times l$ AWG is associated with a wavelength set $\Lambda = \{\lambda_0, \lambda_1, \cdots, \lambda_{|\Lambda|}\}$, where $g$ is the number of inputs, $l$ is the number of outputs, and $|\Lambda| = \max\{g, l\}$. The $g \times l$ AWG is passive wavelength router. Given input $p$ and wavelength $\lambda_i$, the output is uniquely determined:

$$q = [i - p]_{|\Lambda|}, \qquad (1)$$

where $p = 0,1,\cdots, g-1$, $q = 0,1,\cdots, l-1$, and $i = 0,1,\cdots, |\Lambda|$. For example, in Fig. 1(b), wavelength 3 at input 1 reaches output $q = [3 - 1]_6 = 2$.

According to the wavelength routing property (1), each input wavelength channel or output wavelength channel can be labelled by a two-bit address. In particular, wavelength channel $\lambda_i$ at input $x_2$ can be labelled by
$$X = x_2 x_1 = x_2 [i - x_2]_{|\Lambda|} \qquad (2)$$
where $x_2$ and $x_1$ are port bit and wavelength bit respectively. Similarly, wavelength channel $\lambda_k$ at output $z_2$ can be labelled by
$$Z = z_2 z_1 = z_2 [k - z_2]_{|\Lambda|} \qquad (3)$$
As an example, in Fig. 1(b), the wavelength channel $\lambda_3$ at input 1 is labelled as $Y = y_2 y_1 = 1[3-1]_6 = 12$.

With the numbering scheme, it is easy to see that the $2 \times 3$ AWG in Fig. 1(b) also possesses the same connection property as the shuffle network in Fig. 1(a). For example, in Fig. 1(b), input wavelength channel 10 connects with output wavelength channel 01. This conclusion holds in general case, which can be explained as follows. Consider an input channel $X = x_2 x_1$. According to the routing property (1) of AWGs, $x_2 x_1$ connects to the output channel $Z$ with the port bit
$$\left[[x_2 + x_1]_{|\Lambda|} - x_2\right]_{|\Lambda|} = x_1,$$
and the wavelength bit
$$\left[[z_2 + z_1]_{|\Lambda|} - z_1\right]_{|\Lambda|} = z_2.$$
Thus, the output channel address $Z = z_2 z_1 = x_1 x_2$, which agrees with the connection property of shuffle networks.

Though a single AWG can fulfill the function of a shuffle network in principle, it cannot be used to implement a large-scale shuffle network, due to the following three reasons. Firstly, an AWG with large port count will suffer a serious coherent crosstalk [8]. Secondly, the synthesis of AWGs with large port count is very difficult. Thirdly, a large-scale AWG is associated with a large number of wavelengths, which however are precious resource in the optical communication window. Thus, it is necessary to consider the scalability of AWG-based shuffle networks.

## 3. Modular AWG-based Shuffle Network

In this section, we show a large-scale optical shuffle network can also be implemented by a two-stage AWG-based interconnection network, in which the size of each AWG is small.

An example of the two-stage AWG-based interconnection network, denoted as $\mathcal{W}(g, m, n)$, is plotted in Fig. 2(a). In the first stage, there are $gm$ inputs, each of which contains $n$ wavelengths, are evenly divided to $g$ groups. In the second stage, there are $m$ $g \times n$ AWGs. The $b$th port of input group $a$ connects to the $a$th input of the $b$th AWG, where $a = 0, 1, \cdots, g-1$, and $b = 0, 2, \cdots, m-1$. The example shown in Fig. 2(a) is $\mathcal{W}(3,2,3)$, in which the $0^{\text{th}}$ port of input group 1 connects to the $1^{\text{st}}$ input of AWG 0.

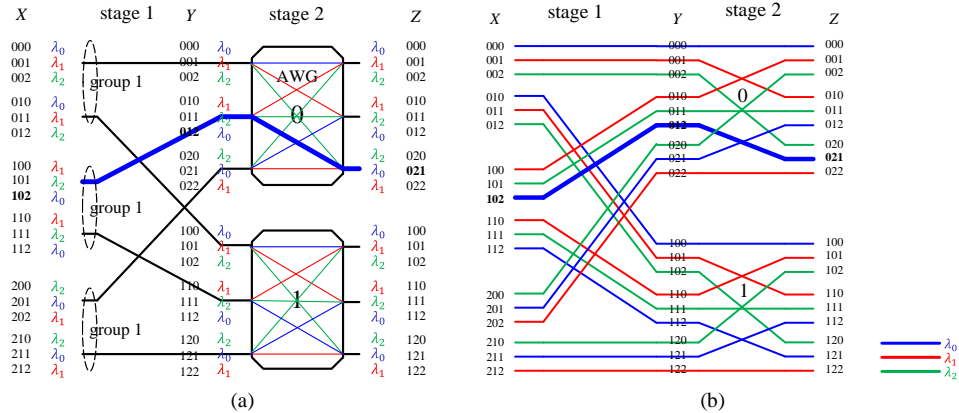

Fig. 2 An example of (a) AWG-based shuffle network $\mathcal{W}(3,2,3)$ and (b) its space representation

According to the wavelength routing property (1), each input wavelength channel or output wavelength channel can be labelled by a three-bit address. In particular, wavelength channel $\lambda_j$ at input $y_2$ of AWG $y_3$, which is referred to as middle channel, can be labelled by
$$Y = y_3 y_2 y_1 = y_3 y_2 [j - y_2]_{|\Lambda|}. \qquad (4)$$
where $y_3$, $y_2$, and $y_1$ are called group bit, port bit, and wavelength bit, respectively. On the contrary, the wavelength of middle channel $Y = y_3 y_2 y_1$ is $\lambda_{[y_1 + y_2]_{|\Lambda|}}$ according to (4). For example, the wavelength index of middle channel 012 is $[y_1 + y_2]_{|\Lambda|} = [2+1]_3 = 0$. Similarly, wavelength channel $\lambda_k$ at input $z_2$ of AWG $z_3$, which is called output channel, can be numbered by
$$Z = z_3 z_2 z_1 = z_3 z_2 [k - z_2]_{|\Lambda|}. \qquad (5)$$

Also, middle channel $Y = y_3 y_2 y_1$ is wavelength $\lambda_{[y_1+y_2]_{|A|}}$ at input $y_2$ of AWG $y_3$, which connects with the same wavelength at $y_3$th port of the $y_2$th input group according to the definition of the two-stage AWG-based network. It follows that wavelength $\lambda_{[y_1+y_2]_{|A|}}$ at $y_3$th port of the $y_2$th input group can be denoted by

$$X = x_3 x_2 x_1 = y_2 y_3 y_1. \tag{6}$$

For example, in Fig. 2(a), wavelength $\lambda_0$ at port $x_2 = 0$ of input group $x_3 = 1$ is labeled as $102$, since it connects with middle channel $Y = 012$, which is wavelength $\lambda_0$ at port $1$ of AWG $0$.

With the help of numbering scheme, we show that $\mathcal{W}(g, m, n)$ is an $N \times N$ optical shuffle network, where $N = gmn$. Fig. 2(b) gives the space representation of the AWG-based interconnection network $\mathcal{W}(3,2,3)$ in Fig. 2(a), where each link in Fig. 2(b) is corresponding to a wavelength channel in Fig. 2(a). From Fig. 2(b), we can see that stage 1 performs a bit shift from input channel $X = x_3 x_2 x_1$ to middle channel $Y = x_2 x_3 x_1$, and stage 2 carries out another bit shift from middle channel $Y = y_3 y_2 y_1$ to output channel $Z = y_3 y_1 y_2$. For example, in Fig. 2(a), the input channel $X = 102$ connects to the middle channel $Y = 012$, and then reaches the output channel $Z = 021$. Thus, $\mathcal{W}(3,2,3)$ exhibits the same connection property with the $18 \times 18$ shuffle network $\mathcal{S}(3,6)$ in Fig. 1(a) and the $2 \times 3$ AWG in Fig. 1(b).

The general case of $N \times N$ two-stage AWG-based shuffle networks $\mathcal{W}(g, m, n)$ is plotted in Fig. 3(a), where $N = gmn$. From Fig.3 (a), it is easy to see that it is allowable to construct a large-scale shuffle network using a set of small-size AWGs. Moreover, when the parameters $g$, $m$, and $n$ vary, there is a tradeoff between the number of wavelengths associated with the network and the number of cables between two stages. To demonstrate such tradeoff, let's consider the case where $g < n$ and $l = mn$ are fixed. When $n$ increases, the size of each AWG and thus the number of wavelengths associated with the network increase. At the same time, the number of cables in the first stage decreases. Especially, when $m = 1$ and $n = l$, the AWG-based shuffle network will change to a $g \times l$ AWG, denoted by $\mathcal{W}(g, 1, l)$. In this case, the number of wavelengths needed by an AWG-based shuffle network is $l$, while the cables in stage 1 disappear. Thus, the two-stage AWG-based shuffle network provides a large design flexibility in practical applications.

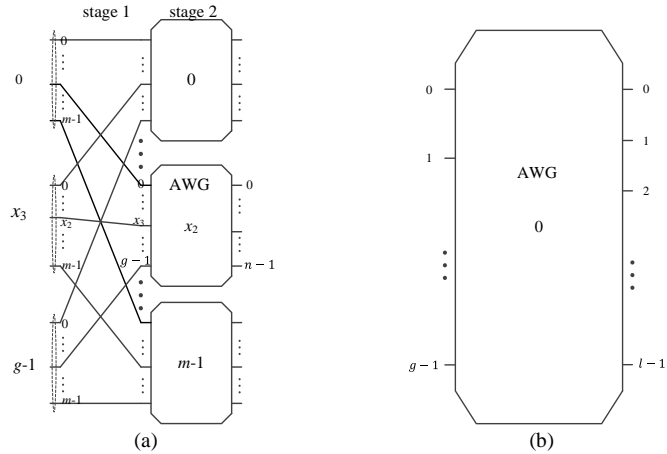

Fig. 3 (a) General Case of AWG-based shuffle network $\mathcal{W}(g, m, n)$ and (b) the extreme case $\mathcal{W}(g, 1, l)$ where $l = mn$ and $g < n$.

## 4. Conclusion

In this paper, we propose a modular AWG-based shuffle network, which is compact, scalable, and easy to implement. Also, the proposed AWG-based shuffle network provides a large design flexibility in practical applications.